\documentclass[showpacs,preprintnumbers,amsmath,amssymb]{revtex4}
\allowdisplaybreaks
\usepackage{graphicx}
\input epsf
\usepackage{hyperref}
\hypersetup{colorlinks=true, citecolor=blue, linkcolor=red}
\newcommand{\sn}{\text{\rm sn}}
\newcommand{\beq}{\begin{equation}}
\newcommand{\beqn}{\begin{equation*}}
\newcommand{\enq}{\end{equation}}
\newcommand{\enqn}{\end{equation*}}
\newcommand{\eb}{{\rm e}}
\newcommand{\R}{{\mathbb R}}
\renewcommand{\Re}{\text{\rm Re}}
\renewcommand{\Im}{\text{\rm Im}}

\begin{document}
\allowdisplaybreaks
\title{A hidden life of Peregrine's soliton: rouge waves in the oceanic depths}
\author{A.A. Yurova}
\email{yurov@freemail.ru}
\affiliation{Baltic Federal University of I. Kant, Theoretical Physics Department, Al.Nevsky St. 14, Kaliningrad 236041, Russia\\}
\affiliation{Kaliningrad State Technical University, Mathematics Department, Sovetsky Av.,1, Kaliningrad, 236000, Russia \\}
\date{\today}
\begin{abstract}
Although the Peregrine-type solutions of the nonlinear Sch\"odinger equation have long been associated mainly with the infamous ``rouge waves'' on the surface of the ocean, they might have a much more interesting role in the oceanic depths; in this article we show that these solutions play an important role in the evolution of the intrathermocline eddies, also known as the ``oceanic lenses''. In particular, we show that the collapse of a lens is determined by the particular generalization of the Peregrine soliton -- so called exultons -- of the nonlinear Schr\"odinger equation. In addition, we introduce a new mathematical method of construction of a vortical filament (a frontal zone of a lens) from a known one by the Darboux transformation.
\end{abstract}


\maketitle

\section{Introduction} \label{sec:Intro}

Since its inception in 1964 \cite{CGT}, the nonlinear Schr\"odinger equation (NLS)
\beq \label{NLS}
i u_t + u_{xx} + 2|u|^2 u = 0,
\enq
has been a subject of an intensive study in the mathematical and theoretical physics. The equation has a wide range of applications starting out from the propagation of the waves on a water surface and ending up with plasma physics and the nonlinear optics. It allows for a plentitude of solutions, such as the ``bright'' and ``dark'' soliton solutions, breathers and even the N-soliton solutions. However, even among this already impressive array, one solution stands on its own and attracts a special attention of both theoreticians and the experimentalists: a famous rouge wave solution. Initially discovered in 1983 by Peregrine \cite{P}, it realized a previously unthought of possibility -- a soliton with both a spatial and temporal localizations:
\beq \label{solNLS}
\mid u\mid^2=A^2+2\frac{p(t)-\xi^2}{(p(t)+\xi^2)^2},
\enq
where
\beqn
p(t)=\frac{1}{4A^2}+(2At)^2,\qquad \xi=x-2at,
\enqn
and $A$ and $a$ are real constants. This unusual behavior led the professional oceanologists to believe that it is this wave that may be at least partially responsible for the famous oceanic ``rouge waves'', a strange kind of a localized wave that appears seemingly out of nowhere in the open seas and then promptly disappears again, causing a considerable damage to those sea vessels that had a misfortune to be on it's way. The Peregrine soliton seemed to be a first clear and mathematically precise model of this devastating phenomena. However, there still remained a concern about the solution \eqref{solNLS}, which, albeit being an exact one, could well have been unique and virtually unobservable. This concern has been addressed in 1992 by Matveev and Salle, who have shown that the Peregrin's soliton is but a member of a family of solutions of NLS that they have dubbed the ``exultons'' \cite{MS}. The name stems from the Latin language (``exilio'' is the verb used when something ``appears suddenly'') and is a tribute to a simple observation: that $|u|^2\to A^2$ as $t \to \pm\infty$. The Peregrin's soliton, for example, is a nonsingular pulse
\beq \label{PerS}
u(x,t)=A \left(-1+\frac{A^{-2} + 4 i t}{p(t) + \xi^2}\right) \eb^{i(a x + (2A^2-a^2)t)}
\enq
which, for all sufficiently large $t > 0$, behaves as
\beqn
\mid u \mid^2 = A^2 + \frac{A^2 - a^2}{2 (A^2+a^2)^2} ~t^{-2} + O(t^{-3}), \qquad t \gg 1.
\enqn
On the other hand, in \cite{MS} it has been shown that there also exist the exulton-like solutions that approach its asymptotic value $t\to\pm\infty$ in an exponential manner:
\beq \label{exul}
\mid u\mid^2=A^2 \left( 1 - \frac{4 h (1 - h^2)\left(h + \cosh P(t)\cos Q(\xi)\right)}{\left(\cosh P(t) + h \cos Q(\xi)\right)^2}\right),
\enq
where the quantities $P(t)$, $Q(\xi)$ and $h$ are defined as
\beqn
P(t) = 4\lambda t \sqrt{A^2-\lambda^2},\qquad
Q(\xi) = \sqrt{A^2-\lambda^2}\xi, \qquad
h = \left|\frac{\lambda}{A}\right| < 1,
\enqn
and $\lambda, ~A$ are real. Moreover, a solution of a similar nature has been procured by Ablowitz and Herbst in a study of a homoclinic structure and a numerically induced chaos for the NLS equation \cite{AH}:
\beq \label{AsolNLS}
\mid u\mid^2 = \frac {a^2 \Big(\mu^2 E(t)^4 + 4 E(t)^3 \mu \cos(\nu x)\cos(2\alpha) + 2 E(t)^2 \big(\mu\cos(4\alpha) + 2 \cos^2(\nu x)\big) + 4 E(t) \cos(\nu x) \cos(2\alpha) + 1\Big)}{(\mu E(t)^2+2\cos(\nu x)E(t)+1)^2},
\enq
where
\beqn
\mu=1/\cos^2\alpha,\qquad
\nu=2a\sin\alpha,\qquad
E(t) = \exp\left(2 a^2 \sin(2 \alpha) t\right),
\enqn
and $\mid u\mid\to |a|=$const as $t\to\pm\infty$.

Thus, by the end of the XX century the theoretical grounds have been well-prepared and everything was ready for an experimental verification. And, after many trials and tribulations, the ``rouge wave'' have finally been detected: first in 2010 in the optical fibers \cite{Po10}; then, a year later -- in the waves generated in the multicomponent plasma \cite{Pp11}; until finally the pulse with the behavior characteristic of a Peregrin soliton has been triumphantly discovered in an experimental water tank \cite{Pw11}. The physical presence of a Peregrine soliton -- a ``rouge wave'' -- has become a fait accompli.

The next question that had to be answered was that of a physical sense of the solutions like \eqref{exul} and \eqref{AsolNLS}. Are they nothing more but a more complicated version of a ``rouge wave'', or do they bear a much deeper significance in a bigger picture of an oceanic dynamics?.. We claim that the latter is true by providing a phenomena, whose behavior perfectly fits the bill. The phenomena in question is the frontal vorticity of isolated eddies, which are also commonly referred to as the intrathermocline eddies (ITE), or simply as ``lenses''.


\section {Intrathermocline  eddies} \label{sec:Eddies}

The contemporary literature dedicated to the phenomenon of the oceanic lenses is rather extensive, stretching from the purely mathematical studies of lenses' theoretical aspects to many a volume of the observational data (see e.g. \cite{DMHS}). The usual definition goes like this: a ``lens'' is an anti-cyclonic eddy that has the following properties:
\begin{itemize}
\item the overall shape of a horizontally oriented biconvex lens;
\item a smallness of the vertical gradients of the inner temperature, salinity and density;
\item the possible inclusions of extraneous waters;
\item a localization mainly in the subsurface and the intermediate water layers.
\end{itemize}

The term ``intrathermocline'' is a tribute to the fact that the lenses are localized deep in the thermocline region (sometimes even in a pycnocline), far from the surface of the ocean. The lenses usually have a thickness of some hundreds of meters, a length of a few kilometers in diameter, and do rotate counter-cyclonically (a cyclonic rotation is actually possible but is significantly less frequent) with the orbital velocities of up to 30 cm/s \cite{KR, AZ}. The lenses are widespread in the oceans -- since 1976 to 1992, for example, more then 200 of lenses have been identified in the Atlantic, Pacific and Indian Oceans \cite{BK1}. They are also one of the most long-lived objects in the ocean -- by some estimates, their life span might stretch for more than a decade \cite{MR}, which is made even more astounding by the fact that some of the lenses have been observed at 6--7 thousands of kilometers from the place of their origin \cite{BK2}.

So, what brings to life these unusual objects?.. The most popular theory of a lens' generation associates it with intrusive water flows extending along the intrinsic horizontal isopycnal surfaces. Because of a non-zero viscosity, the velocity of an intrusive tongue will differ in the center and in its outermost areas, thus turning the area confined within an arbitrary ``liquid contour'' (chosen inside of a tongue) into a time-dependent one. On the other hand, by the Bjerknes circulation theorem, the resulting vorticity will be a subject of change under the Coriolis force. Various numerical experiments demonstrate that when an initial vorticity is non-homogeneous, the evolution of an intrusive flow results in the separation of an eddy and, thus, a subsequent creation of a new lens.

A new-born lens will only be stable if the mixing effects are sufficiently small; this also implies that the vorticity will be focused on a frontal, almost circular, zone of a lens. This frontal vorticity is often called a {\bf vortex filament}, since a ratio of vertical to horizontal characteristic scales for the oceanic lenses does not exceed $10^{-3}$. The reason for this has been ascertained in the course of laboratory experiments with density lenses in a rotating fluid \cite{KN, Z}. These experiments have shown that early in the formation process of the lens the horizontal pressure gradient flattens it in the vertical direction (the so-called collapse of lens), thus transforming the frontal vortex filament into an essentially flat curve. Moreover, because the horizontal projection of a lens is almost circular, the filament with time degrades not just into a flat curve, but a flat curve of a constant curvature, i.e. a circle. The importance of this observation will become apparent once we turn out attention to an interpretation of the solutions \eqref{solNLS}--\eqref{AsolNLS}.

Before we conclude this section, we would like to discuss two important experimental results that have been independently obtained in the works of Kostyanoy-Shapiro \cite{KS1,KS2} and Kitamura-Nagata \cite{KN}. Both groups studied the behavior of a rotating lens formed via the intrusion process. In both experiments shutting down the source of an intrusion flow resulted in a viscous Ekman relaxation of a lens, manifesting itself in a slow gradual expansion of radius of the lens and a decrease in its thickness (hence, the term ``collapse''). However, the exact way these processes unfolded differed between the two experiments. In the Kitamura-Nagata case the surface radius of a lens has been expanding as $R \approx \sqrt{t}$ while the thickness behaved as $H = H_0 \exp(-c t)$, where $H_0$ is an initial thickness of a lens and $c>0$ depended on the viscosity and the density of the intrusive flow. Quite different results have been obtained by the Kostyanoy-Shapiro group. Their experiments have shown the radius of a lens increasing as $R \approx t^d$ with $d = 0.25 \pm 0.07$, while the thickness decreased not exponentially but as $H \approx 1/t^{c}$ with $c > 0$. An explanation to this almost scandalous discrepancy has been proposed by A. Zatsepin in the article \cite{Z}. His idea was that the two groups have observed the lenses in two {\em different} phases of their lives. In particular, the Kitamura-Nagata group studied the initial (so-called inertial) phase of lens formation, characterized by an exponential relaxation of such major characteristics as thickness of the lens and the curvature of its front. This regime cannot not hold forever, so it soon gives way to the polynomial regime of collapse -- in particular, changing the law that governs the behavior of the lens thickness to $H \approx t^{-1/2}$, just as observed by Kostyanoy and Shapiro. Eventually, a viscosity will put an end to a decrease of $H$ altogether, leading to a last, viscous stage of the evolution of the lens with a proper circular shape and a constant thickness $H$ \cite{KS2}. It is important to note that, although the last two phases are relatively well-studied, it is the first (initial or inertial) phase that proves to be most difficult for both the observations and the mathematical modeling. The problem gets aggravated by a requirement for a smooth transition of this regime into a second stage of the collapse.

The main goal of the current article is to offer a way of creating models of evolution of the most significant part of an intrathermocline eddy -- its frontal vorticity, with the specific requirement: that the internal characteristics of the eddy (namely, the curvature and the torsion) would be governed by an exponential law that recedes in time, giving way for a polynomial dependence that, in turn, as $t \to +\infty$ yields a vortex filament of a proper circular shape with constant curvature and torsion. Most surprisingly, this goal ends up being directly related to the task of establishing a {\em superposition} of Peregrine soliton with an exponentially behaving exulton, as will be demonstrated in Sec. \ref{sec:Collapse}. But before we get there, we are going to address the pivotal question of why and how does NLS equation \eqref{NLS} relates to the parameters of an arbitrary vortex filament.


\section{The NLS and the collapse of a vortex filament} \label{sec:Collapse}

In the paper \cite{H} Hasimoto has shown that when a vorticity is confined within a long narrow tube (a vortex filament) with its characteristics being constant across the tube, the evolution of the curvature and torsion of a filament will be described by the NLS equation \eqref{NLS}. The result goes like follows.

Let $k=k(x,t)$ be the curvature of a vortex filament, $\varkappa=\varkappa(x,t)$ -- the filament's torsion, and $x$ -- a natural parameter along it. If the velocity of the vortex filament only have a binormal component, then a new function $u=u(x,t)$, defined as
\beq \label{u}
u(x,t)\equiv 2k(x,t)\exp\left\{i\left(\int_0^x dx'\, \varkappa(x',t) + \int_0^t dt' F(t')\right)\right\},
\enq
will be a solution to the NLS equation \eqref{NLS} (where $F(t)$ is some arbitrary function). Thus, if the orthogonal component $\vec v_{_N}$ of the velocity $\vec v$ is invariant throughout the vortex filament, simply switching to a reference frame moving with the velocity $\vec v_{_N}$ and perusing the corresponding nonlinear Schr\"odinger equation would allow one to predict the further evolution of a shape of the vortex filament.

To be more precise, let $u(x,t) = u_r(x,t) + i u_i(x,t)$, where $u_r$ and $u_i$ are two real-valued functions. Introducing two new functions $r = \mid u \mid = \sqrt{u_r^2 + u_i^2}$ and $q = u_i / u_r$, we may rewrite the equation \eqref{u} as the system
\beq \label{kkap}
\begin{split}
k(x,t) & = r(x,t) d/2 \\
\varkappa(x,t) & = \partial_x \left(\arctan q(x,t) \right).
\end{split}
\enq
The system \eqref{kkap} has a number of consequences. In particular, the zeroes of the curvature of vortex filament apparently correspond to zeroes of the solution $u$, while the zeroes of the torsion are the products of either the zeroes of $u_r$ or of a local linear dependance between the imaginary and real parts of $u$. Interestingly, the assumption that torsion is identical to zero everywhere might only be satisfied by a purely real (or purely imaginary) time-independent solution of \eqref{NLS} of the form
\beqn
\mid u_c(x) \mid = \mid a ~\sn (a x; -1) \mid,
\enqn
where $\sn(x;m)$ is a standard Jacobi elliptic function and $a \in \R$.

Another simple but interesting application of the formula \eqref{kkap} is associated with the traveling wave solution of the NLS equation \eqref{NLS}
\beq \label{trav}
v(x,t) = A \exp \{i\left( a x + (2A^2-a^2)t \right)\},
\enq
where $a, A \in \R$. Due to a specific form of solution \eqref{trav}, the parameter $q(x,t)=\tan (a x + (2A^2-a^2)t)$ which implies that the corresponding vortex filament solution has both constant torsion $\varkappa = a$ and curvature $k = A/2$. In fact, it is easy to show that this is the only solution of \eqref{NLS} with such properties.

Returning back to our case and using the formulas \eqref{kkap} for the Peregrine soliton \eqref{solNLS} (assuming for the sake of simplicity that $a=0$) we obtain the following formula:
\beqn
\varkappa=\frac {128 A^4xt}{256A^8t^4+128A^6(xt)^2+16A^4(x^4+10t^2)-24A^2x^2+9}.
\enqn
The similar calculations can be performed for solutions \eqref{exul} and \eqref{AsolNLS} (we do not show the calculations here only due to the cumbersomeness of the resulting formulas). It should now be clear that the solutions \eqref{solNLS}--\eqref{AsolNLS} suits the purpose of simulation of the lenses' collapse rather well. Indeed, as $t\to+\infty$, $\varkappa\to 0$, and $k\to =$const, to the solutions \eqref{solNLS}--\eqref{AsolNLS}. Let us note here that all these solutions permits an introduction of new parameters -- in particular, the ones that make sure the asymptotic limit of torsion is a non-zero number, albeit an infinitesimal one. In these cases the filament will degrade into a quasy-flat (but not absolutely flat) curve.

The solutions \eqref{exul}, \eqref{AsolNLS} provide us with two examples of the closed vortex filaments, that can be associated with a frontal vorticity of a collapsing eddy. The solutions \eqref{solNLS} and \eqref{exul} were obtained via Darboux transformations \cite{MS} whereas the solution \eqref{AsolNLS} has been procured with the aid of Hirota method \cite{AH}. It is possible to construct a superposition of both exponential and rational solutions \eqref{solNLS} and \eqref{exul}. To do this, however, one should first derive the nonlinear superposition formulae for the NLS equation via the Darboux transformation.

The LA-pair for NLS has the form
\beq \label{LA}
\hat L\Psi=\Lambda \Psi,\qquad \Psi_t=\hat A\Psi,
\enq
where $\hat L$ and $\hat A$ are two matrix-valued ($2\times 2$) differential operators (w.r.t. the variable $x$), of the first and second order correspondingly \cite{MS};
\beqn
\Psi=\left(\begin{array}{cc}
\psi_1&\psi_2\\
-\psi_2^*&\psi_1^*,
\end{array}\right),
\qquad
\Lambda=\left(\begin{array}{cc}
l&0\\
0&l^*
\end{array}\right),
\enqn
and $l$ is a complex spectral parameter.

Let $\Psi_{1,2}$ be two solutions of the system  \eqref{LA} for the respective values of the spectral parameter $l_{1,2}$ and let $u$ be a fixed solution of the NLS. Define the matrix functions $\tau_k=\Psi_k\Lambda_k\Psi_k^{-1}$, $k=1,\,2$. Then, the LA-pair equations are invariant under the Darboux transformation,
\beq \label{Darboux}
\Psi\to\Psi[1]=\Psi\Lambda-\tau\Psi,\qquad
u\to u[1]=u-2(l-l^*)\frac{\psi_1\psi_2^*}{\mid\psi_1\mid^2+\mid\psi_2\mid^2}.
\enq
In particular, the solutions \eqref{solNLS} and \eqref{exul} have been explicitly derived by \eqref{Darboux} under the assumptions $u\equiv u_0 = A\exp(2iA^2t)$ and $a=0$.

In order to construct the nonlinear superposition formulae one should make use of the fact that the two sequential Darboux transformations can be rewritten as a product of two {\em commuting} operators:
\beqn
u[1,2]=u[2,1]=u_3.
\enqn
Using this fact, we will, after some straightforward calculations, get a sought after nonlinear superposition formula for the solution \eqref{solNLS} ($u_1$) and \eqref{exul} ($u_2$):
\beqn
u_3=\frac {1}{2}(u_1+u_2)+u_{int},
\enqn
where
\beqn
u_{int}=\frac {(g^*_1-g^*_2)^2(u_1+u_2-2u_0) + (u_1-u_2)^2(u_1+u_2-2u_0)^*-2(g_1^2-g_2^2)^*(u_1-u_2)} {\mid g_1-g_2\mid^2+\mid u_1-u_2\mid^2},
\enqn
with
\beqn
g_k = \frac {i}{\mid u_0\mid^2-\mid u_k\mid^2} \left(u_0 \partial_x w^*_k-u^*_k\partial_x w_k\right), \qquad w_k \equiv u_k - u_0.
\enqn

It is possible to choose the parameters in such a way that the exponential solution $u_2$ will dominate at $t$ close to zero, and the rational solution $u_1$ will start dominating as $t\to\infty$.

\section{Constructing the brand new vortex filaments}\label{sec:Darboux}

We have seen in the previous section that the NLS equation can be indispensable if one wishes to construct the vortex filaments through its curvature and torsion. What remains unresolved is the problem of recovery	of the curve's shape from a given curvature and torsion. In this section we will  discuss this problem and, in addition, we present a method of construction of a vortical filament from a known one, using the Darboux transformation.

Let us denote a vector tangential to the vortical filament by $\vec v$; $\vec n$ and $\vec b$ will denote a normal and binormal vectors, respectively. These values are interrelated through the Frene formulas. Our goal now is to find a shape of a corresponding vortical filament by means of parametrization
\beqn
\vec r(x,t)=\int_a^x dx'\vec v(x',t),
\enqn
where $\vec r(x,t)$ is a radius-vector of the points belonging to the curve. The problem then reduces to finding the tangent vector as the function of a natural parameter and time. This, of course, requires solving three Ricatti equations \cite{St},
\beq \label{Ricat}
\partial_x\psi_{_k}+ik(x,t)\psi_{_k}-\frac {i\varkappa(x,t)}{2}(1-\psi_{_k}^2)=0,\qquad
k=1,2,3.
\enq
Here
\beqn
\psi_{_k}\equiv \frac {v_{_k}+in_{_k}}{1-b_{_k}},
\enqn
$v_{_k}$, $n_{_k}$ and $b_{_k}$  are the components of the appropriate vectors. The knowledge of $\psi$ allows one to recover the exact form of $\vec v$ (although it will contain at least one arbitrary parameter, associated with a way one chooses the coordinate system). The task of determining the  shapes of the vortex filaments from the solutions of NLS can be performed by the by the means of the Sym formula, and it has been extensively studied in such papers as \cite{Si}, \cite{C1}, \cite{C2}. In this article we decided to concentrate instead on deriving the explicit expressions for the curvature and torsion. However, there is still one interesting point we would like to discuss here first. After linearization of \eqref{Ricat} via the substitution $\psi = -2i\partial_x\ln\Phi/\varkappa$, one obtains the equation of the second order
\beq \label{linRic}
\Phi_{xx} + \left[ik(x,t)-\partial_x\ln{\varkappa(x,t)}\right] \Phi_x+(\varkappa(x,t)/2)^2\Phi=0.
\enq
This equation is not easy to integrate. If, however, one does manage to find a full set of exact linearly independent solutions $\{\Phi_{_{1,2}}\}$ for at least some values of $k(x,t)$ and $\varkappa(x,t)$:
\beqn
\Phi_{_2}=\frac {\Phi_{_1}}{\sqrt{\varkappa}}\exp\left\{\frac i2\int dx'k\right\}\cdot
\int dx'\frac {\varkappa}{\Phi_{_1}^2}\exp\left\{i\int dx''k\right\},
\enqn
then it would be possible to convey a systematic procedure of construction of new integrable potentials, quite analogous to the Darboux transformation \cite{MS, D, C}. Indeed, let  $\Phi_{_1}$ and $\Phi_{_2}$ be two solutions of \eqref{linRic}. A direct computation shows that the function
\beqn
\Phi_{_2}[1]=S\left(\Phi_1\partial_x \Phi_{_2} - \Phi_2\partial_x \Phi_{_1}\right)/\Phi_1,
\enqn
satisfies \eqref{linRic} with the new potentials (we omit the arguments of functions for the sake of brevity):
\beq \label{kapps}
\begin{array}{l}
\displaystyle
{\varkappa[1]^2 =-{\varkappa^2}-4\left(up+u_x+
\frac{S_{xx}+S_x(u+p)}{S}-\partial_x(\ln\Phi_{_1})
\left(2u+p+2\partial_x\ln(S/\Phi_{_1})\right)\right),} \\
{}\\
\displaystyle
{k[1]=k-ip+i\partial_x\ln(\varkappa[1]/\varkappa),}
\end{array}
\enq
where $u\equiv -ik+(\ln\varkappa)_x$. The functions $S(x,t)$ and $p(x,t)$ should be chosen in a way that complies with $\varkappa[1], k[1]$ being real-valued. In particular, the absence of an imaginary part of $k[1]$ implies that the real part of $p$, which we denote by $p_{_R}$, equals
\beq
\label{pr}
\Re~ p = p_{_R} = (\ln \varkappa[1]/\varkappa)_x,
\enq
while the imaginary part $\Im ~p = p_{_I}$ remains for the time being an arbitrary quantity. Substituting \eqref{pr} into the first formula of \eqref{kapps} and introducing a new function $\chi\equiv\ln\varkappa[1]$, we obtain a new nonlinear differential equation for $\chi$,
\beq \label{chi}
F_1\chi_x+F_2=\exp{\{2\chi\}}
\enq
and the condition of the real-valuedness of the functions $F_{1,2}$ is the necessary condition for the real-valuedness of $\chi$. It is convenient at this point to represent the functions $\Phi_{_1}$ and $S$ as exponential functions
\beqn
\Phi_{_1}=A\exp\{i\alpha\},\qquad S=B\exp\{i\beta\}.
\enqn
The condition $\Im ~F_1 = 0$ leads to the formula
\beqn
\beta=\alpha+\int dl\,k,
\enqn
which, together with the similar condition for $F_2$, allows one to calculate $p_{_I}$. As the result, we obtain the following law of the curvature transformation,
\beq \label{curv}
\displaystyle
{k[1]=k+\frac {(2k+\alpha_x)_x+k\left(\ln{B}/{\varkappa}\right)_x
+\alpha_x\left(\ln{(\varkappa B^4)}/{A^2}\right)_x}
{\left(\ln{\varkappa}/{(AB)}\right)_x}.}
\enq

In order to find a similar formula for the torsion, one should solve the nonlinear equation \eqref{chi}. By decreasing its order, one can reduce it to the Ricatti equation, which can be represented along the same lines as \eqref{linRic}. If one will manage to integrate it (by finding such a function $B$ that allows a maximal simplification of the potential involved), the problem will be solved. The second solution of the transformed equation \eqref{linRic} might be obtained by an application of the formula \eqref{kapps}. In other words, once a curvature and torsion that allows for a reconstruction of a shape of a curve (that is, finding the general solution of equation \eqref{linRic}) have been found, we can consequently construct an infinite number of filaments of the same property. Let us note that, in contrast to the routine method of construction of the integrable potentials by the means of Darboux transformation, we have to use not one but two {\bf integrable} linear equations of the second order.

By way of example let's consider a curve of the constant curvature and torsion. The solution of equation \eqref{linRic} have the form
\beq \label{newsol}
\Phi_{_1}=\cos\left(\frac {\sqrt{k^2+\varkappa^2}\,x}{2}\right)
\exp\left\{-\frac {ikx}{2}\right\}.
\enq
Then
\beq \label{aDef}
\begin{split}
\alpha & =-\frac {\pi}{2} \text{sign} \left(\sin\left(\frac {kl}{2}\right)\cos\left(\frac {\sqrt{k^2+\varkappa^2}\,l}{2}\right)\right)+
\pi\left[\frac {kl}{2\pi}\right]+\frac {\pi-kl}{2}\\
A & =2 \cos\left(\frac {\sqrt{k^2+\varkappa^2}\,l}{2}\right),
\end{split}
\enq
and the braces in the second term indicate that the number therein should be lowered down to a closest integer. Using \eqref{aDef} we end up with the formula for a new curvature
\beqn
k[1]=\frac {4kB_x \cos\left({\sqrt{k^2+\varkappa^2}\,x}/{2}\right)} {2kB_x \cos\left( {\sqrt{k^2+\varkappa^2}\,x}/{2}\right) - \sqrt{k^2+\varkappa^2}B\sin\left({\sqrt{k^2+\varkappa^2}\,x}/{2}\right)},
\enqn
which can be associated with a frontal vorticity of a lens undergoing a collapse in an oscillating regime. Computation of coefficients $F_1$ and $F_2$ leads to rather cumbersome expressions, which we will omit here.

The transformation described above, which we'll further refer to as the Darboux transformation, allows for more iterations. To see this, let us rewrite the analogue of equation \eqref{Ricat} after $N$ iterations:
\beq \label{RicatN}
\Phi[N]''+P[N]\Phi[N]'+Q[N]\Phi[N]=0,
\enq
where by $P[N]$ and $Q[N]$ we have defined the following quantities
\beq
\label{PQN}
P[N]\equiv ik[N]-(\ln\varkappa[N])',\qquad Q[N]\equiv \frac {\varkappa [N]^2}{4}.
\enq

For our next step we will require exactly $(N+1)$ special solutions of \eqref{LA}: $\{\Phi_{_1},\,\Phi_{_2},...,\,
\Phi_{_N},\,\Phi_{_{N+1}}\equiv\Phi\}$, $\lambda_{_i}\ne \lambda_{j}$,
$\lambda_{_{N+1}}=0$. It is easy to see that the N-times iterated function $\Phi[N]$ can be written as
\beq \label{PhiN}
\Phi[N]=S(\Phi^{(N)}+\sum_{k=1}^{N}\,a_k\Phi^{(N-k)}),\qquad \Phi^{(k)}\equiv
\frac {\partial^k\Phi}{\partial l^k}.
\enq
Substituting \eqref{PhiN} into \eqref{RicatN} yields
\beq \label{PhiN2}
\begin{split}
S\Phi^{(N+2)}+&(2S'+SP[N])\Phi^{(N+1)}+(S''+S'P[N]+SQ[N])\Phi^{(N)}+\\
&\sum_{k=1}^{N}\,[\left((Sa_k)''+(Sa_k)'P[N]+Sa_kQ[N]\right)\phi^{(N-k)}+\\
&\left(2(Sa_k)'+Sa_kP[N]\right)\Phi^{(N-k+1)}+Sa_k\Phi^{(N-k+2)}]=0.
\end{split}
\enq
Next, we solve \eqref{RicatN} for $\Phi''$ and subsequently differentiate it $N$ times. This produces
\beq \label{PhiN22}
\Phi^{(N+2)} = -\sum_{m=0}^{N}C_N^m \left( P^{(N-m)}\Phi^{(m+1)}+Q^{(N-m)}\Phi^{(m)} \right),
\enq
where $C_N^m=\frac {N!}{(N-m)!m!}$; substituting \eqref{PhiN22} into \eqref{PhiN2} and assuming the linear independence of the various derivatives of $\Phi$, we end up with the following expressions for $P[N]$ and $Q[N]$:
\beq
\label{QP}
\begin{split}
Q[N] &= Q+NP'-\frac {(S'-2Sa_1)'}{S}-\left(\frac {S'}{S}+a_1\right)P[N]-a_2,\\
P[N] &= P-\frac {2S'}{S}-a_1.
\end{split}
\enq

Let $S=\exp(f_R+i f_I)$. Recalling from \eqref{PQN} what $P$ and $Q$ amounts for, the second equation from the system \eqref{QP} yields
\beqn
k[N]=k-2i(\ln\frac {\varkappa[N]}{\varkappa})'+2i(\ln S)'+ia_1.
\enqn
Imposing a physically sound requirement that $k[N] \in \R$ results in new formulas for both a torsion
\beq
\label{vark}
(\ln\frac {\varkappa[N]}{\varkappa})'=2f_R'+a_{1,R},
\enq
and a curvature
\beq
\label{k}
k[N]=k-2f_I'-a_{1,I},
\enq
where it is important to keep in mind that the $f_I$ is an arbitrary function.

Now we turn our attention to the first equation of the system \eqref{QP} and require the torsion to be a real-valued function. This implies that the imaginary part of this equation is identically equal to zero. What follows is the relationship for $f_R$:
\beq
\label{fR}
f_R'=-\frac {a_{1,R}(2a_{1,I}+f_I'-k)+(a_{1,R}+f_I')(\ln\varkappa)'-(a_{2,I}+
2a_{1,I}-Nk+f_I')'}{a_{1,I}+2f_I'-k}.
\enq
Upon the substitution of \eqref{fR} into the first equation of \eqref{QP} we end up with
\beq
\label{vark1}
\frac {\varkappa[N]^2}{4}=\frac {\varkappa^2}{4}-N(\ln\varkappa)'+
\frac {Z_1}{Z_2},
\enq
where
\beqn
\begin{split}
Z_1 = &a_{1,R}^2(3a_{1,I}^2+3a_{1,I}(f_I'-k)+3(f_I')^2-3kf_I'+k^2)+a_{1,R}(2a_{1,I}^2(\ln\varkappa)'-\\
&a_{1,I}(3a_{2,I}+6a_{1,I}'-(3N+1)k'+2(3(f_I')^2-(\ln\varkappa)'f_I'))+ka_{2,I}+a_{1,I}'(3f_I'+k)-\\
&k'(f_I'+Nk)+2(kf_I''+(\ln\varkappa)'f_I'(f_I'-k)))-a_{1,I}^4+a_{1,I}^3(3k-5f_I')-a_{1,I}^2(a_{2,f_I}+\\
&9(f_I')^2-11kf_I'+3k^2)-a_{1,I}(2a_{2,f_I}(2f_I'-k)+(\ln\varkappa)'a_{2,I}+a_{1,R}'(3f_I'-k)+2a_{1,I}''+\\
&2(\ln\varkappa)'a_{1,I}'+a_{2,I}'-Nk''-(N+1)(\ln\varkappa)'k'+f_I'''+2(\ln\varkappa)'f_I''+8(f_I')^3-14k(f_I')^2+\\
&f_I'(7k^2+((\ln\varkappa)')^2)-k(k^2+((\ln\varkappa)')^2))-a_{2,f_I}(4(f_I')^2-4kf_I'+k^2)+a_{2,I}^2+a_{2,I}(5a_{1,I}'-\\
&(2N+1)k'+4f_I''-(\ln\varkappa)'k)-a_{1,R}'(6(f_I')^2-5kf_I'+k^2)+2a_{1,I}''(k-2f_I')+6(a_{1,I}')^2-\\
&a_{1,I}'((5N+2)k'+(\ln\varkappa)'(3k-f_I'))+a_{2,I}'(k-2f_I')+Nk''(2f_I'-k)+N(N+1)(k')^2+\\
&k'((\ln\varkappa)'(f_I'+Nk)-9f_I''-(4N+1)f_I'')(k-2f_I')+3(f_I'')^2-2(\ln\varkappa)'kf_I''-f_I'(4(f_I')^3-\\
&8k(f_I')^2+f_I'(5k^2+((\ln\varkappa)')^2)-k(k^2+((\ln\varkappa)')^2)),\\
Z_2 = &a_{1,I}+2f_I'-k.
\end{split}
\enqn
The compatibility condition of \eqref{vark} and \eqref{vark1} has the form
\beq
\label{compat}
\frac {1}{2}\left\{\ln\left(\varkappa^2-4N(\ln\varkappa)'+4\frac {Z_1}{Z_2}\right)
\right\}'=(\ln\varkappa+2f_R)'+a_{1,R}.
\enq
Choosing $f_I$ that satisfies \eqref{compat} leads to the following formula for the torsion:
\beqn
\varkappa[N]=\sqrt{\frac {\left(\varkappa^2-8N(\ln\varkappa)'+
8\frac {Z_1}{Z_2}\right)'}{2[(\ln\varkappa +2f_R)'+a_{1,R}]}}.
\enqn

What remains unknown here are the coefficients $a_{1,2}$. In order to find them, let us first notice that the $N$-th iterations of each of the functions $\{\Phi_{_1},\,\Phi_{_2},...,\,\Phi_{_N}\}$ are identically equal to zero. Thus, we have a system of $N$ linear equations:
\beqn
\Phi_j^{(N)}+\sum_{k=1}^{N}\,a_k\Phi_j^{(N-k)}=0,\qquad j=1,...,N.
\enqn
Hence, the coefficients in question can be found out using the standard Cramer's formulas:
\beqn
a_1=-\frac {D_1}{D},\qquad a_2=\frac {D_2}{D},
\enqn
and
\beqn
D=\vmatrix
\Phi_1^{(N-1)}&\Phi_1^{(N-2)}&\Phi_1^{(N-3)}&...&\Phi_1\\
\Phi_2^{(N-1)}&\Phi_2^{(N-2)}&\Phi_2^{(N-3)}&...&\Phi_2\\
.\\
.\\
.\\
\Phi_N^{(N-1)}&\Phi_N^{(N-2)}&\Phi_N^{(N-3)}&...&\Phi_N
\endvmatrix
\enqn
\beqn
D_1=\vmatrix
\Phi_1^{(N)}&\Phi_1^{(N-2)}&\Phi_1^{(N-3)}&...&\Phi_1\\
\Phi_2^{(N)}&\Phi_2^{(N-2)}&\Phi_2^{(N-3)}&...&\Phi_2\\
.\\
.\\
.\\
\Phi_N^{(N)}&\Phi_N^{(N-2)}&\Phi_N^{(N-3)}&...&\Phi_N
\endvmatrix
\enqn
\beqn
D_2=\vmatrix
\Phi_1^{(N)}&\Phi_1^{(N-1)}&\Phi_1^{(N-3)}&...&\Phi_1\\
\Phi_2^{(N)}&\Phi_2^{(N-1)}&\Phi_2^{(N-3)}&...&\Phi_2\\
.\\
.\\
.\\
\Phi_N^{(N)}&\Phi_N^{(N-1)}&\Phi_N^{(N-3)}&...&\Phi_N
\endvmatrix
\enqn

\section{Conclusion} \label{sec:Conclusion}

In this article we have demonstrated that although the famous Peregrine solution of the Nonlinear Schr\"odinger equation (NLS) is usually considered a paragon of a menace that are the rouge waves, it might play a mich subtler role in the vortical dynamics of the world Ocean, for it describes the behavior of a frontal vorticity of such a widespread phenomena as the intrathermocline eddies. This phenomena, also known as the ocean lenses, forms as a by-product of an intrusion of a ``tongue'' of foreign waters into a water mass of different density/salinity (see Sec. \ref{sec:Eddies}).

A new-born horizontally-oriented eddy undergoes a number of changes, such as the vertical flattening and smoothening in the edges, and eventually acquires the shape remarkably similar to that of an optical lens -- hence the name. Due to the effects of viscosity, the vorticity of the lens focuses in the frontal area, which, thanks to the aforementioned flattening, turns into what is called a vortical filament. The process of conversion of an initial eddy into a well-formed lens with a frontal vortical filament of the nearly-constant torsion and curvature is called the collapse of the lens, and it can be broken into three phases. The first, inertial phase, is characterized by exponential behavior of its characteristics (curvature, torsion, height). During the second, intermediate one, the regime is switched from the exponential to polynomial. The last, viscous phase, is when the characteristics of a lens reach their asymptotic values. We have shown that, using the Hasimota method (see Sec. \ref{sec:Collapse}), each one of these regimes might be associated with the designated Peregrine-type solitons of the NLS. Interestingly, the Peregrine soliton itself corresponds only to the second and the last stages of the collapse of a lens. The first one, being exponential by nature, requires the exponential generalization of the Peregrine soliton, which can be obtained by the means of a Darboux transformation performed for a periodic (in time) solution of NLS. The overall solution thus ends up being a linear superposition of the two, with the exponential one dominating for small values of $t$ and the polynomial solution taking over for large $t$. Finally, in Sec. \ref{sec:Darboux} we have demonstrated that there exist an algorithm akin to a Darboux transformation, that allows to construct new curvature and torsion functions from the previously known ones. The resulting functions are compatible with NLS and thus also corresponds to some vortex filament. By way of example, we have constructed a curvature function, which can be associated with a frontal vorticity of a lens undergoing an initial stage of a collapse in an oscillating regime.
\newline
\newline
{\bf Acknowledgements.} I would like to thank Artyom Yurov for many stimulation discussions and Valerian Yurov for his help in making this article a fait accompli. My gratitude also goes to professor Cie\'sli\'nski for his kind suggestions.



\begin{thebibliography}{xxxxxx}
\section*{BIBLIOGRAPHY}

\bibitem[CGT]{CGT} R. Y. Chiao, E. Garmire, C. H. Townes, ``Self-Trapping of Optical Beams'', {\em Phys. Rev. Lett.} {\bf 13}, 479--482 (1964)

\bibitem[P]{P} D. H. Peregrine, ``Water waves, nonlinear Schr\"odinger equations and their solutions'', {\em J. Austral. Math. Soc. B} {\bf 25}, 16–-43 (1983)

\bibitem[Po10]{Po10} B. Kibler, J. Fatome, C. Finot, G. Millot, F. Dias, G. Genty, N. Akhmediev and J. M. Dudley, ``The Peregrine soliton in nonlinear fibre optics'', {\em Nature Physics} {\bf 6}:10, 790–-795 (2010)

\bibitem[Pp11]{Pp11} H. Bailung, S. K. Sharma, Y. Nakamura, ``Observation of Peregrine solitons in a multicomponent plasma with negative ion'', {\em Phys. Rev. Lett.} {\bf 107}, 255005:1--4 (2011)

\bibitem[Pw11]{Pw11} A. Chabchoub, N. P. Hoffmann, N. Akhmediev, ``Rogue wave observation in a water wave tank'', {\em Phys. Rev. Lett.} {\bf 106}, 204502:1--4 (2011)

\bibitem[MS]{MS} V. B. Matveev and M. A. Salle, ``Darboux Transformation and Solitons'', Berlin--Heidelberg: Springer Verlag (1991)

\bibitem[AH]{AH} M. J. Ablowitz and B. M. Herbst, ``On Homoclinic Structure and Numerically Induced Chaos for the Nonlinear Schr\"odinger Equation'', {\em SIAM J. Appl. Math.} {\bf 50}, 2, 339--351 (1990)

\bibitem[DMHS]{DMHS} J. P. Dugan, R. P. Mied, P. C. Mignerey, and A. F. Shuetsz, ``Compact, intrathermocline eddies in the Sargasso Sea'', {\em J. Geophys. Res.}  {\bf 87}, C1, 385--393 (1982)

\bibitem[KR]{KR} A. G. Kostyanoy and V. G. Rodionov, ``The Coastal Upwelling Zones as the Source of The Intrathermocline Eddies' Formation'', {\em Intrathermocline Eddies in the Ocean}, Academy of sciences of the USSR: P.P. Shirshov institute of oceanology, Moscow, 50--55 (1986)

\bibitem[AZ]{AZ} L. Armi and W. Zenk, ``Large Lenses of Highly Saline Mediterranean Water'', {\em J. Phys. Oceanogr.} {\bf 14}, 1560--1576 (1984)

\bibitem[BK1]{BK1} I. M. Belkin, A. G. Kostyanoy, ``Intrathermocline eddies in the ocean and their regional differences'', {\em Coherent structures and self-organization of oceanic flows}, M.: Nauka Publications, 112--126 (1992)

\bibitem[MR]{MR} S. B. McDowell, R. T. Rossby, ``Meditorranean water: An intence mesoscale eddy off the Bahamas'', {\em Science} {\bf 302}, N. 4372, 1085--1087 (1978)

\bibitem[BK2]{BK2} I. M. Belkin, A. G. Kostyanoy, ``Lenses from the mediterranean water in the Northern Atlantic'', {\em Hydrophysical studies under the ``Mesopoligon'' program}, M.: Nauka Publications (1988)

\bibitem[KS1]{KS1} A. G. Kostyanoy, G. I. Shapiro, ``Evolution and structure of an intrathermocline eddy'', {\em Proceed. of Acad. of Sci. of USSR} {\bf 22: 10}, 1098--1105 (1986)

\bibitem[KS2]{KS2} A. G. Kostyanoy, G. I. Shapiro, ``The theoretical and laboratory modeling of the mesoscale oceanic eddies'', {\em Sea Hydrophys. Journal} {\bf 5}, 14--21 (1985)

\bibitem[KN]{KN} Yo. Kitamura and Yu. Nagata, ``Behavior of fresh water injected at the surface of a uniformly rotating ocean'', {\em J. Oceanogr. Soc. Japan} {\bf 39}, 89--100 (1983)

\bibitem[Z]{Z} A. G. Zatsepin, ``Laboratory Experiments with Density Lenses in a Rotating Fluid'', {\em Intrathermocline Eddies in the Ocean}, Academy of sciences of the USSR: P.P. Shirshov institute of oceanology, Moscow, 62--70 (1986)

\bibitem[H]{H} J. H. Hasimoto, ``A soliton on  a vortex filament'', {\em J. Fluid Mech.} {\bf 51}, 477--485 (1972)

\bibitem[St]{St} D. J. Struik, ``Lectures on classical differential geometry'', 2nd Edition, Dover Publications (1988)

\bibitem[Si]{Si} A. Sym, ``Soliton surfaces  II.  Geometric Unification of Solvable Nonlinearities'', {\em Lett. Nuovo Cim.} {\bf 36}, 307--312 (1983)

\bibitem[C1]{C1} J. Cie\'sli\'nski, ``Two solitons on a thin vortex filament'', {\em Phys. Lett. A} {\bf 171}, 323--326 (1992)

\bibitem[C2]{C2} J. L. Cie\'sli\'nski, ``Why the Phase Shifts for Solitons on a Vortex Filament Are So Large: A Theoretical Explanation'', {\em Phys. Rev. Lett.} {\bf 94 (13)}, 134503: 1--4 (2005)

\bibitem[D]{D} J. G. Darboux, ``Sur une proposition relative aux \'equations lin\'eares'', {\em Comptes Rendus Acad. Sci.} {\bf 94}, 1456--1459 (1882)

\bibitem[C]{C} M. M. Crum, ``Associated Sturm-Liouville Systems'', {\em Quart. J. Math. Oxford (2)} {\bf 6}, 121--127 (1955)


\end{thebibliography}
\end{document}